\documentclass[]{spie}  
\usepackage[]{graphicx}
\usepackage{amssymb}
\usepackage{hyperref}

\title{Characterization of deformable mirrors for the MagAO-X project} 

\author{Kyle Van Gorkom\supit{a,b}, Kelsey L. Miller\supit{a,b}, Jared R. Males\supit{b}, Olivier Guyon\supit{a,b,c}, Alexander T. Rodack\supit{a,b}, Jennifer Lumbres\supit{a,b}, Justin M. Knight\supit{a,b}
\skiplinehalf
\supit{a}College of Optical Sciences, University of Arizona, 1630 E. University Blvd., Tucson, AZ 85719, USA
 \skiplinehalf 
\supit{b}Steward Observatory, University of Arizona, 933 North Cherry Avenue, Tucson, AZ 85719, USA
 \skiplinehalf 
\supit{c}Subaru Telescope, National Astronomical Observatory of Japan, 650 North A'ohoku Place, Hilo, HI 96720, USA
}

\begin{document} 
  \maketitle


\begin{abstract}
The MagAO-X instrument is an upgrade of the Magellan AO system that will introduce extreme adaptive optics capabilities for high-contrast imaging at visible and near-infrared wavelengths. A central component of this system is a 2040-actuator microelectromechanical (MEMS) deformable mirror (DM) from Boston Micromachines Corp.\ (BMC) that will operate at 3.63 kHz for high-order wavefront control. Two additional DMs from ALPAO will perform low-order and non-common-path science-arm wavefront correction. The accuracy of the wavefront correction is limited by our ability to command these DMs to a desired shape, which requires a careful characterization of each DM surface. We have developed a characterization pipeline that uses a Zygo Verifire Interferometer to measure the surface response and a Karhunen-Lo\`eve transform to remove noise from our measurements.  We present our progress in the characterization process and the results of our pipeline applied to an ALPAO DM97 and a BMC Kilo-DM, demonstrating the ability to drive the DMs to a flat of $\lesssim$ 2nm and $\lesssim$ 4nm RMS in our beam footprint on the University of Arizona Wavefront Control (UAWFC) testbed.
\end{abstract}



\section{INTRODUCTION}
\label{sec:intro}
MagAO-X is an extreme adaptive optics (ExAO) system for the 6.5m Magellan Clay telescope designed for high-contrast imaging (HCI) at visible and near-infrared wavelengths. MagAO-X will deliver a Strehl ratio of $\gtrsim 0.7$ at H$\alpha$ (0.656$\mathrm{\mu m}$), 14-30 mas resolution, and $10^{-4}$ contrast from $\sim$1 to 10 $\lambda$/D~\cite{males, malesspie}. A vector Apodizing Phase Plate (vAPP)\cite{snik} coronagraph will perform the starlight suppression for HCI. Two deformable mirrors will be used for wavefront control in a woofer-tweeter scheme, with a third dedicated to non-common-path correction (NCP). The system woofer, a high-stroke 11x11 97-actuator deformable mirror (DM) from ALPAO, will provide low-order wavefront correction while the tweeter, a 50x50 2040-actuator DM from BMC, simultaneously provides high-order wavefront correction. A pyramid wavefront sensor (PyWFS) operating at 3.63 kHz will drive these DMs. The third DM, a second ALPAO DM97, will perform the NCP correction in concert with low-order and focal-plane wavefront-sensing in the coronagraph arm\cite{miller}.
\begin{figure}[t]
   \begin{center}
   \begin{tabular}{c}
   \includegraphics[height=9cm]{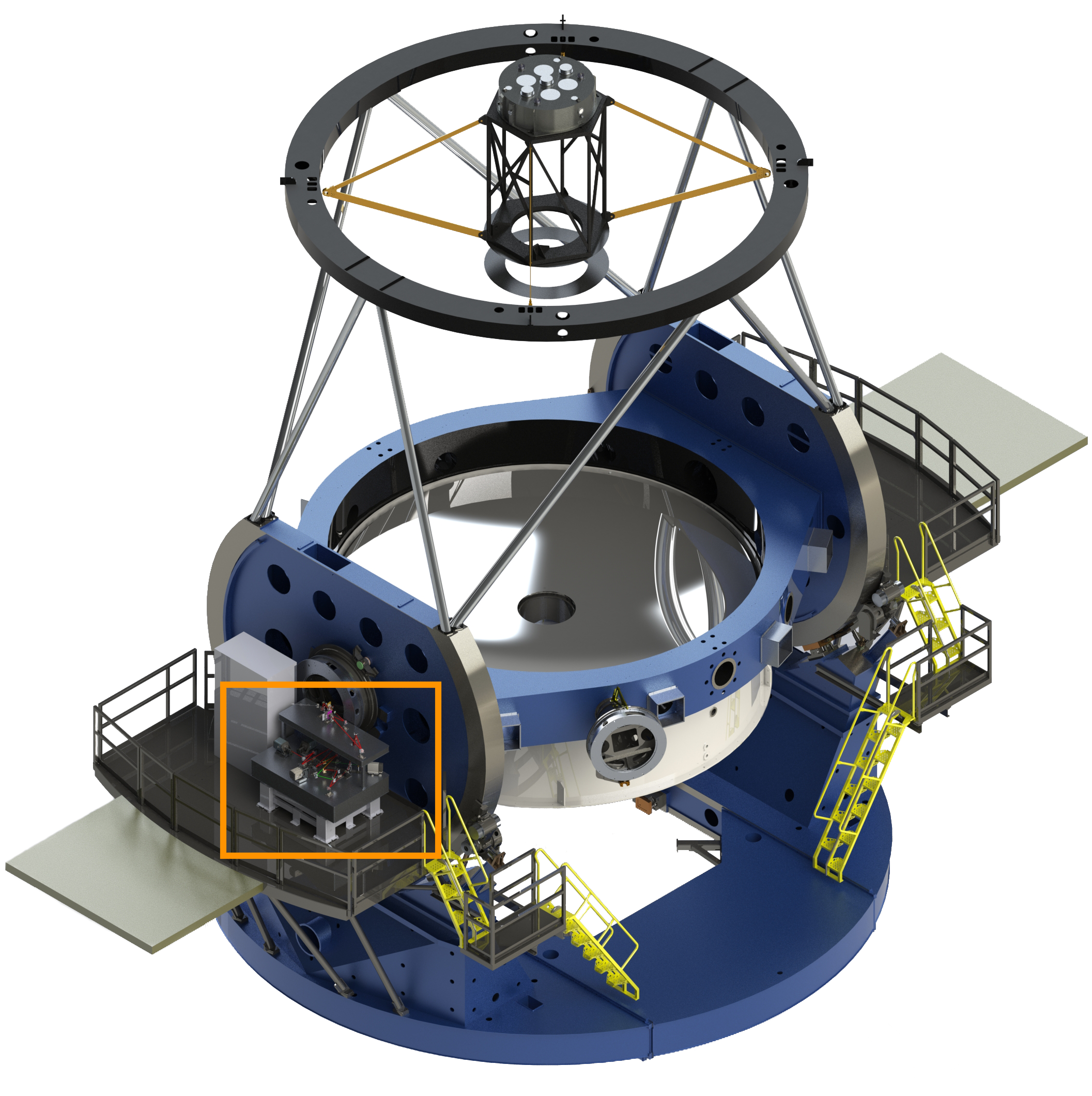}
   \end{tabular}
   \end{center}
   \caption[example] 
   { \label{fig:magtel} A rendering of the Magellan Clay telescope, with the location of the MagAO-X instrument highlighted.}
\end{figure} 
\begin{figure}[h]
   \begin{center}
   \begin{tabular}{c}
   \includegraphics[height=8cm]{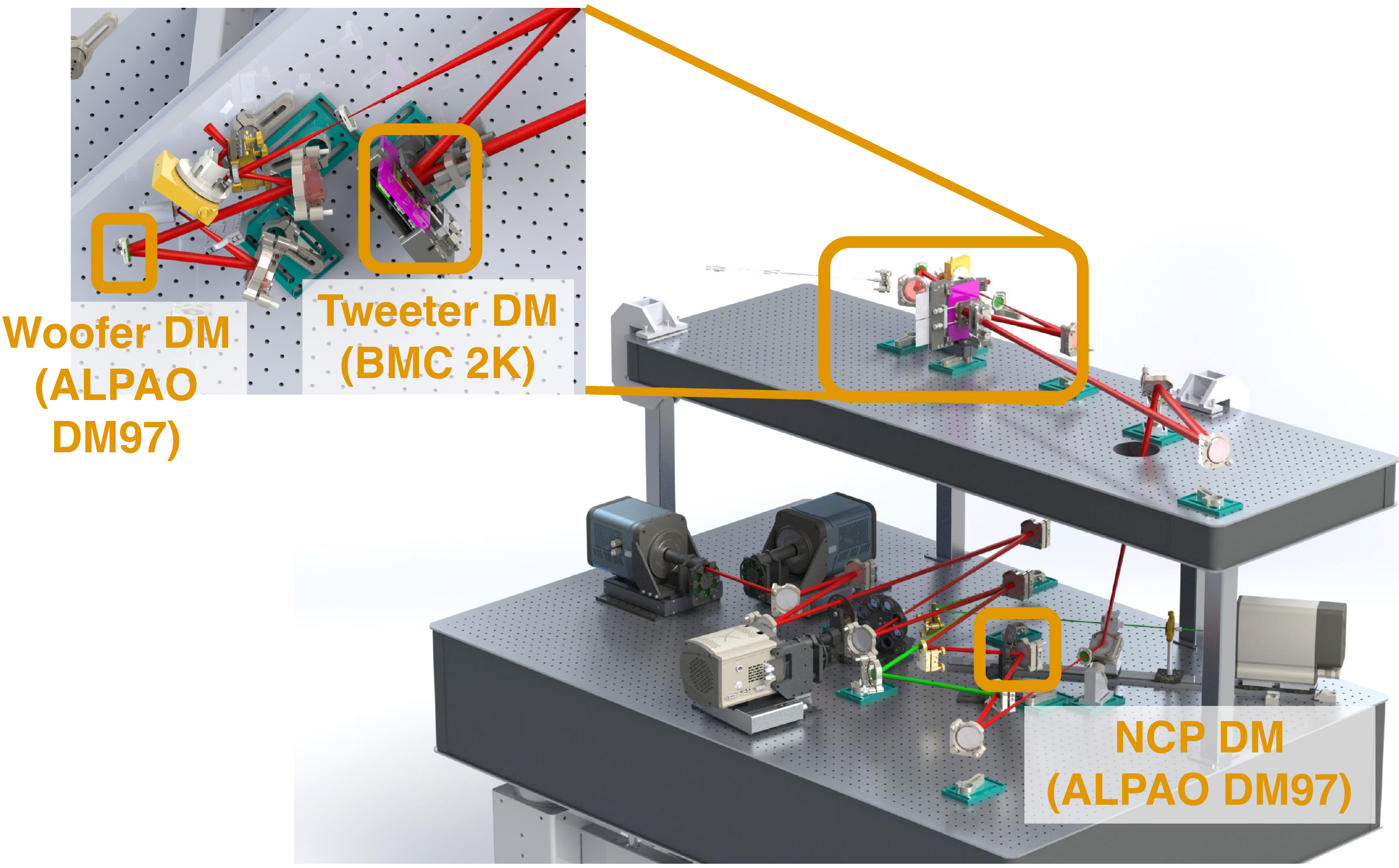}
   \end{tabular}
   \end{center}
   \caption[example] 
   { \label{fig:magtel} A rendering of the MagAO-X bench showing the locations of the deformable mirrors in the optical train. See Close \textit{et al.} 2018 for details of the optomechanical design.~\cite{lairdspie}}
\end{figure} 

Characterization of the DM surfaces, which is necessary for accurate wavefront control, is underway on the UAWFC testbed. To facilitate this process, we have developed a characterization pipeline implemented in Python that uses a Zygo Verifire Interferometer to individually measure the membrane deformation from each actuator (the influence function or IF), which can be combined to create a command matrix that accounts for the overlap of the IFs. To minimize the presence of noise and other undesirable environmental effects which quickly accumulate in this matrix, we project each measured IF onto a Karhunen-Lo\`eve (KL) basis constructed from a library of relaxed DM surfaces and subtract the resulting KL image from the measured IF. This process is described in Sec. \ref{sec:pipeline}. In Sec. \ref{sec:char}, we present the results of our pipeline applied to a 1020-actuator BMC DM and ALPAO DM97. Finally, in Sec. \ref{sec:thefuture}, we discuss future improvements to the pipeline and ongoing work.

\section{CHARACTERIZATION PIPELINE}
\label{sec:pipeline}

\begin{figure}[t]
   \begin{center}
   \begin{tabular}{c}
   \includegraphics[height=7cm]{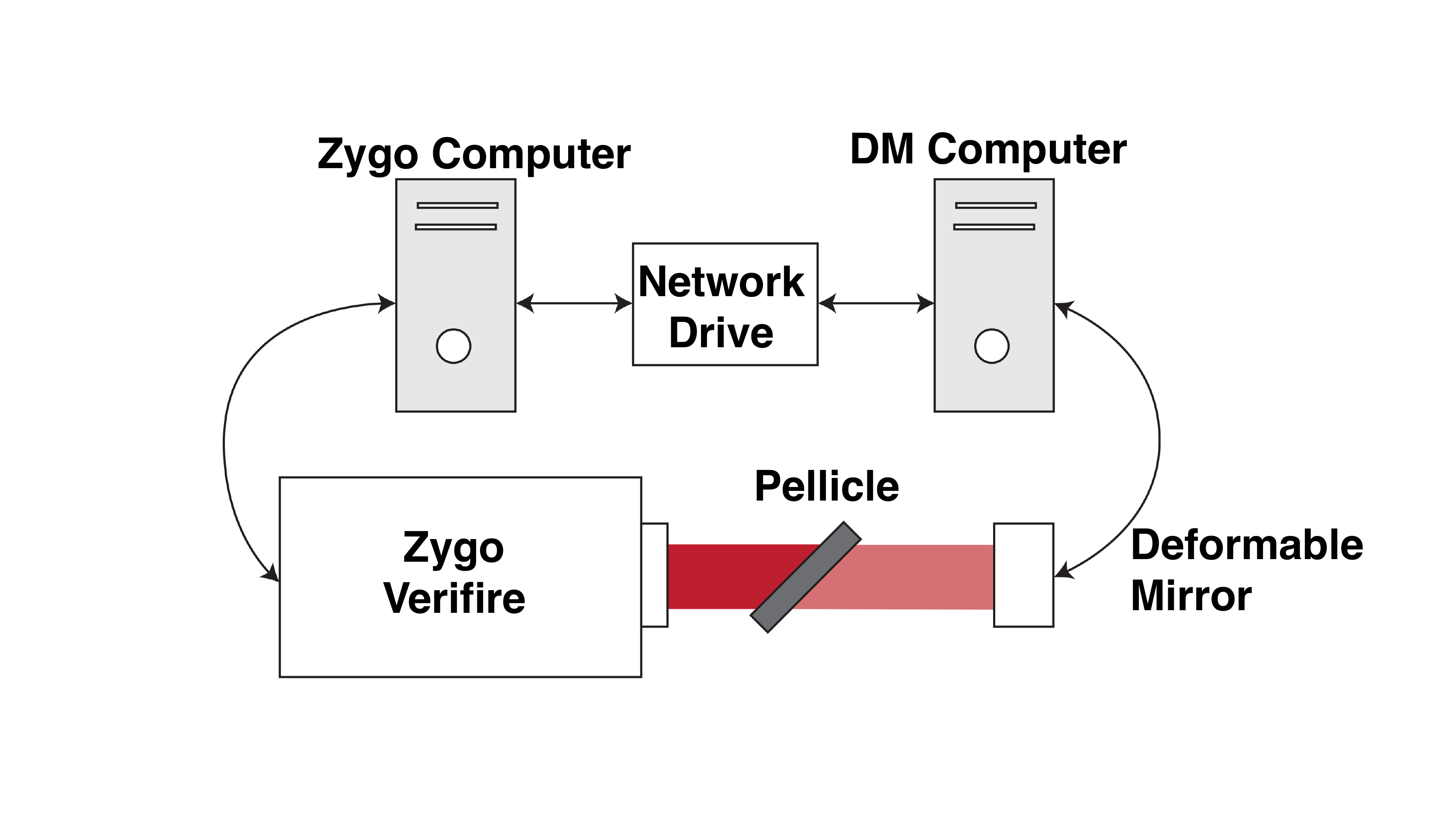}
   \end{tabular}
   \end{center}
   \caption[example] 
   { \label{fig:setup} The optical and hardware setup for DM characterization on the UA wavefront control testbed. The interferometer and DM under characterization are controlled in Python and synchronized across multiple servers through a shared drive.}
\end{figure} 

Before the deformable mirrors can be integrated into the MagAO-X bench, we characterize them on the UA Wavefront Control Testbed. We align each DM to a Zygo Verifire (Fizeau) interferometer to take high-precision measurements of the DM surface. In the current setup, a transmission flat with a 4$\verb+"+$ collimated beam must be paired with a pellicle to maximize fringe contrast. The pellicle imprints additional and unknown content on the DM measurements. We recently acquired a $\lambda/50$ high-reflectance transmission flat from Zygo that will obviate the need for the pellicle and improve future DM measurements.

To expedite the measurement process, we wrote a Python library that automates the interferometer measurements and DM commands and synchronizes the process across multiple servers via a shared network drive (Fig. \ref{fig:setup}). Zygo provides a Python API for measurement acquisition, basic post-processing, and writing to disk. Each DM has a custom API, which we interfaced with the Compute and Control for Adaptive Optics (CACAO) package for low-latency DM control~\cite{guyon}. For any desired characterization procedure, the code accepts a list of DM command inputs and iterates over the list in a command-measure loop. The code is available on github at \url{https://github.com/magao-x/zygo-automation}.

\begin{figure}[h]
   \begin{center}
   \begin{tabular}{c}
   \includegraphics[height=5cm]{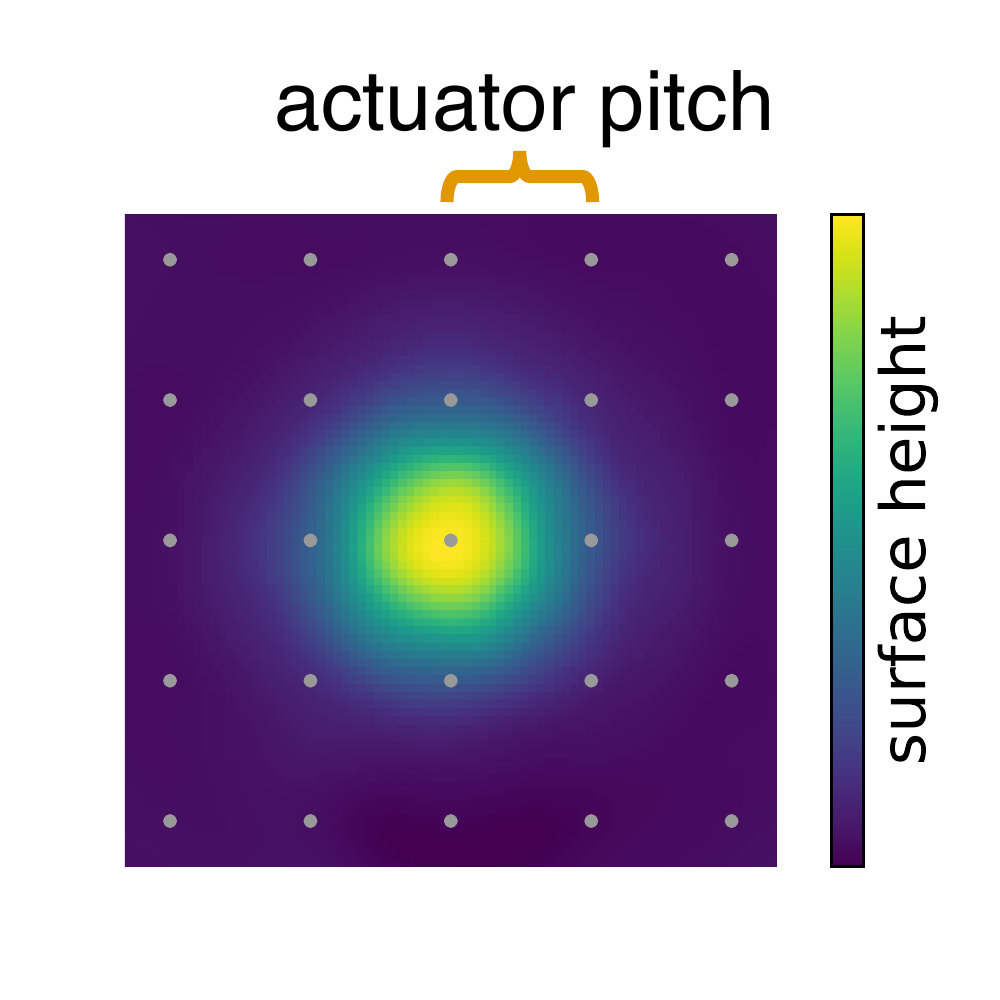}
   \end{tabular}
   \end{center}
   \caption[example] 
   { \label{fig:influence} A representative influence function.}
\end{figure} 

The influence function (IF) of a DM describes how the surface deforms in response to the motion of each actuator (Fig. \ref{fig:influence}). Neglecting inter-actuator coupling, an arbitrary shape on the DM can be approximated as a linear combination of influence functions~\cite{blain}:
\begin{equation}
\vec{s} = F \vec{a}
\end{equation}
where $\vec{s}$ is a vector describing the surface, $\vec{a}$ is a vector of actuator input strokes (which function as weights on the IFs), and $F$ is a matrix describing each actuator's influence function.
If the influence function for each actuator is known, we can find the DM inputs required to produce a desired shape by projecting the shape onto the pseudo-inverse of the influence matrix:
\begin{equation}
\label{eqn:2}
\vec{a} = F^\dagger \vec{s}
\end{equation}

\begin{figure}[t]
   \begin{center}
   \begin{tabular}{c}
   \includegraphics[height=9cm]{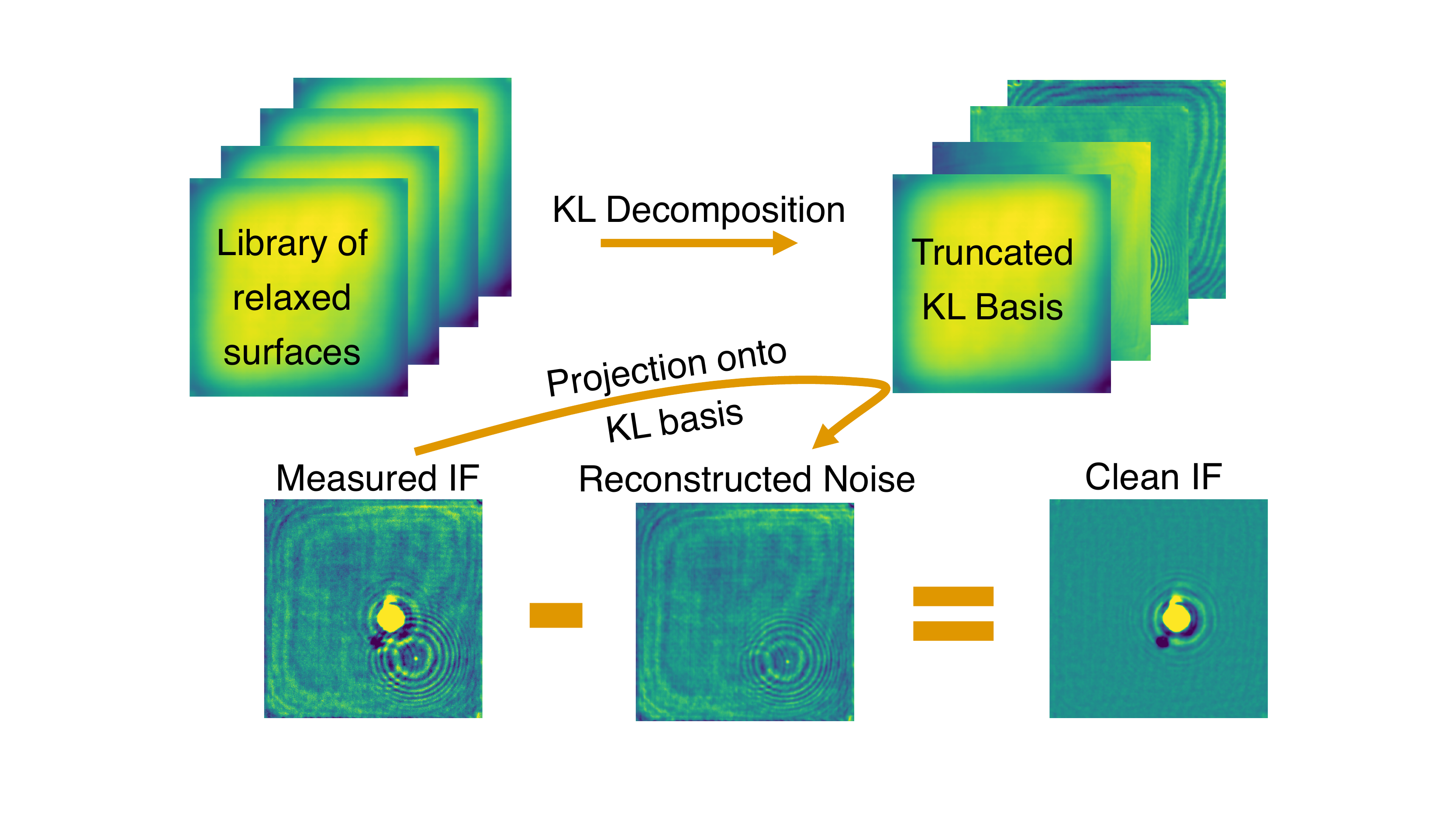}
   \end{tabular}
   \end{center}
   \caption[example] 
   { \label{fig:klip} Influence functions can be cleaned up by projecting the measurement onto a Karhunen-Lo\`eve basis created from a library of unactuated surface measurements. Common sources of noise are dust fringes, phase extraction errors, and low-level surface deformation as environmental conditions change in the lab.}
\end{figure} 

Measuring each influence function precisely is crucial for accurate control of the DM using equation \ref{eqn:2}. Measurement errors, such as mechanical disturbances on the testbed and uncontrollable environmental factors in the lab, tend to accumulate in the influence matrix $F$; to mitigate these errors we use the Karhunen-Loeve Image Projection (KLIP)~\cite{soummer} algorithm to remove rigid body, static surface, and dynamic effects from $F$. While collecting measurements of each actuator IF, we interleave measurements of the un-actuated, or relaxed, DM surface. These relaxed surfaces compose a library of measurements with environmental conditions that are representative of those present during the IF measurements. We then perform a KL decomposition on the library of relaxed DM surface images to generate a basis set of image features that are uncorrelated with the corresponding IF. We project each measured IF onto this basis to reconstruct a noisy surface that can be removed from the IF. Fig. \ref{fig:klip} describes this algorithm pictorially. We then average multiple processed IFs to reduce any variability among individual measurements.

\section{DEFORMABLE MIRROR STATUS AND CHARACTERIZATION}
\label{sec:char}

We are in the process of finalizing the procurement of the DMs for MagAO-X. An ALPAO DM-97 for low-order wavefront correction was delivered in 2017 and a preliminary characterization following the procedure above has been performed. The 2K tweeter DM has been characterized by BMC and will be delivered during the summer of 2018, after which we expect to perform an in-house characterization. A second ALPAO DM-97 for the NCP coronagraph arm will be ordered shortly.

\subsection{BMC 2K}
\label{sec:bmc}

\begin{figure}[t]
   \begin{center}
   \begin{tabular}{c c}
   \includegraphics[height=3cm]{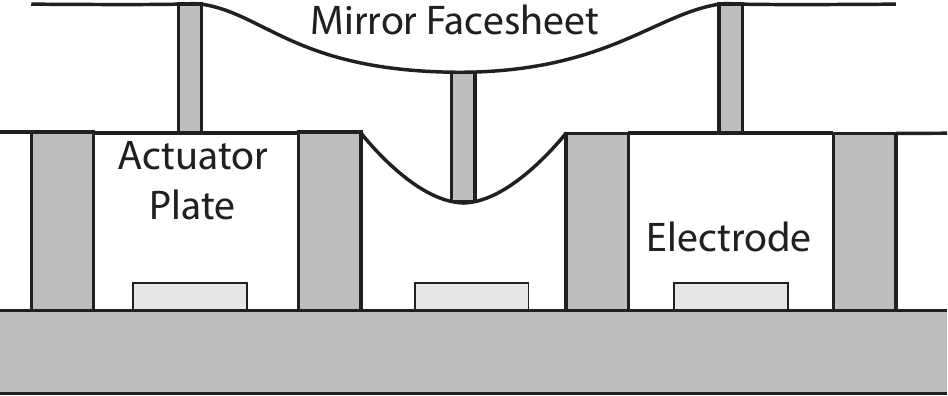}
   \raisebox{-0.2\height}{\includegraphics[height=4cm]{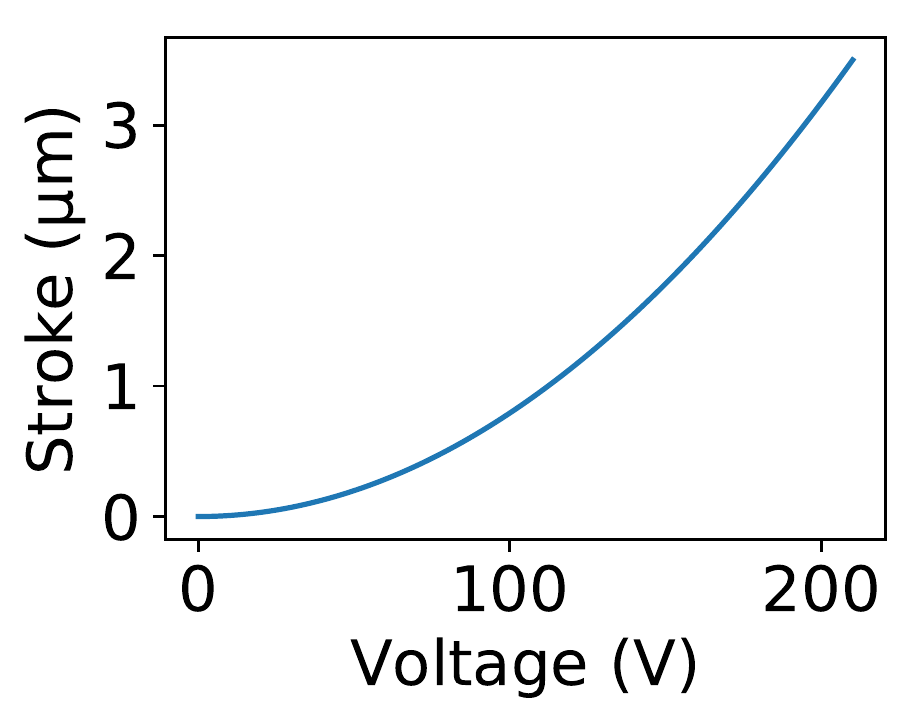}}
   \end{tabular}
   \end{center}
   \caption[example] 
   { \label{fig:bmc} Left: Schematic of the BMC design. The actuator plate is pulled down by an electric field when a voltage is applied to the actuator electrode. Right: The stroke of a single actuator has an approximately quadratic relationship with the applied voltage.}
\end{figure} 

\begin{figure}[b]
   \begin{center}
   \begin{tabular}{c c}
   \includegraphics[height=5.5cm]{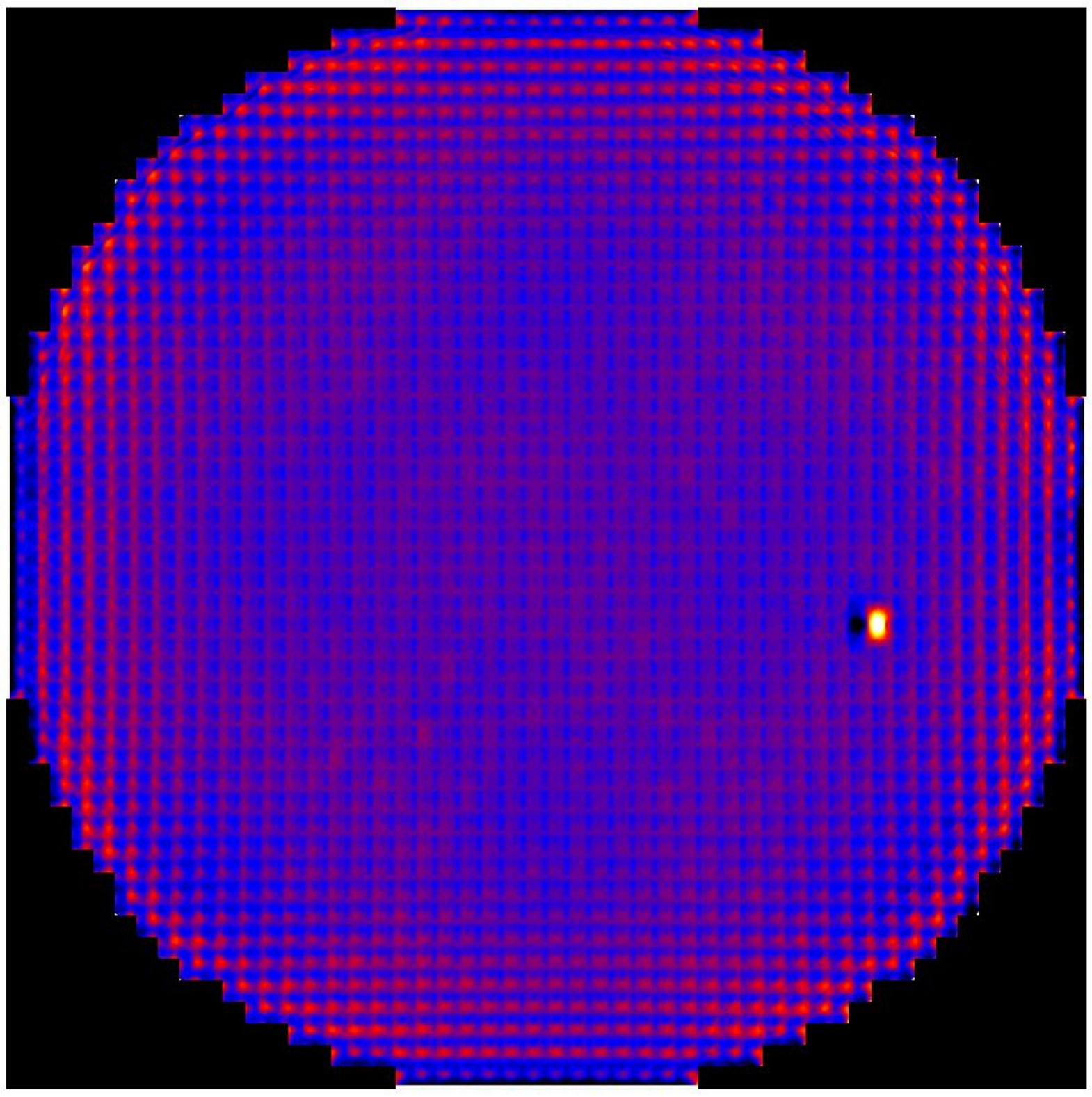}
   \includegraphics[height=5.5cm]{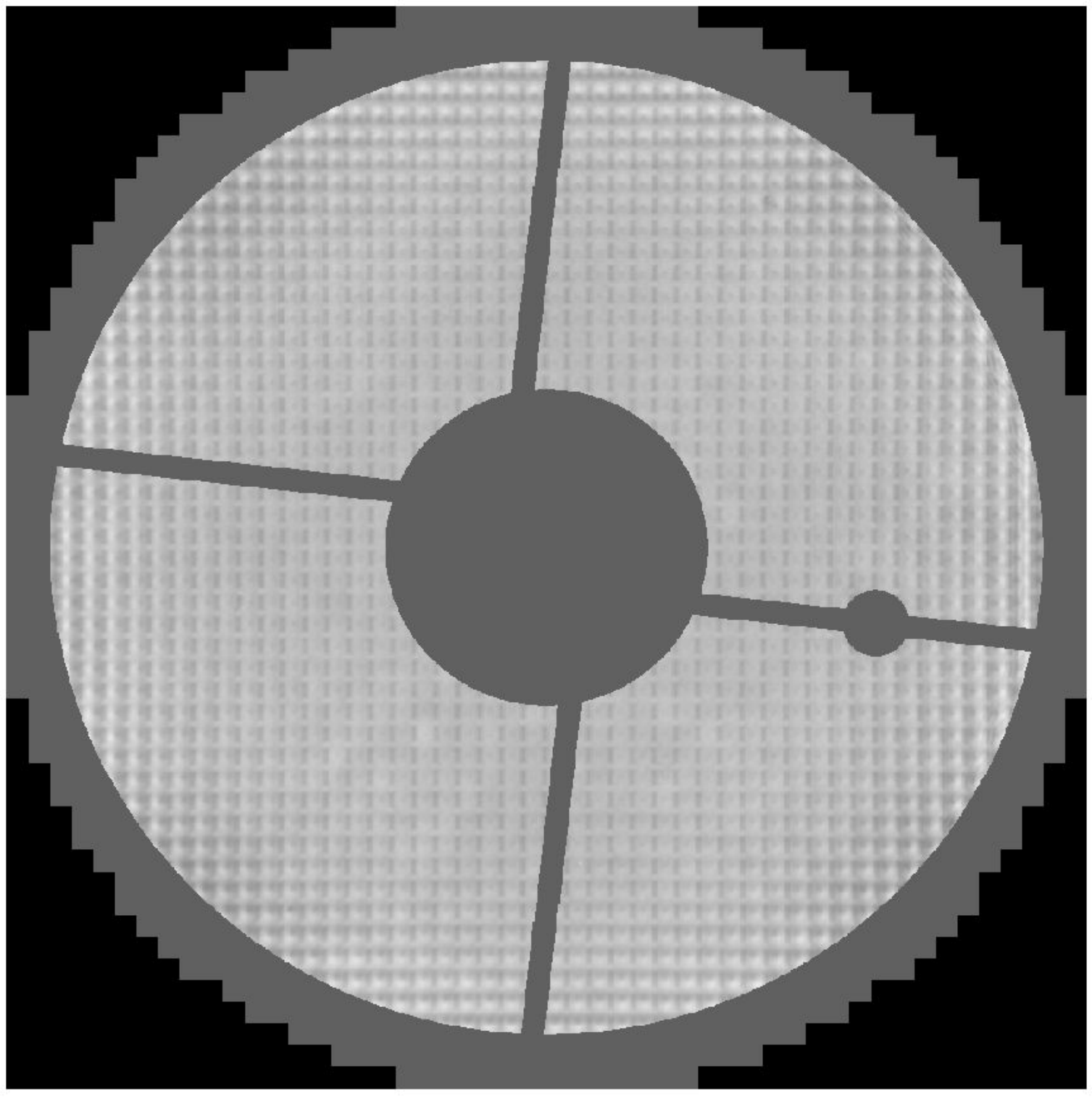}
   \end{tabular}
   \end{center}
   \caption[example]
   { \label{fig:bmc_2k} Left: Best flat on the BMC 2K (16.7nm RMS). Right: The same flat with the MagAO-X coronagraphic pupil with additional actuator mask projected onto it (11.3nm RMS).}
\end{figure} 

The 50x50 2K DM from BMC deforms a continuous facesheet via an electrostatic force between an electrode and actuator plate, resulting in an approximately quadratic relationship between voltage and stroke (Fig. \ref{fig:bmc}). A unique calibration of the voltage-stroke curve for each actuator will be determined in the lab. The specifications of our device are:
\begin{itemize}
\item 2040 actuators
\item 19.6mm circular aperture
\item 3.5$\mu$m stroke
\item 400$\mu$m pitch
\item AR-coated, 6$^\circ$-wedge protective window
\end{itemize}
BMC's characterization of the 2K DM found one significant defect and a 16.7nm RMS flat over the 19.6mm diameter. Over the MagAO-X coronagraph pupil rotated to mask the defect with the spider arms, the surface error reduces to 11.3nm RMS. We plan to confirm these reported values on our testbed upon taking delivery of the DM this summer.

\begin{figure}[t]
   \begin{center}
   \begin{tabular}{c c}
   \includegraphics[height=6cm]{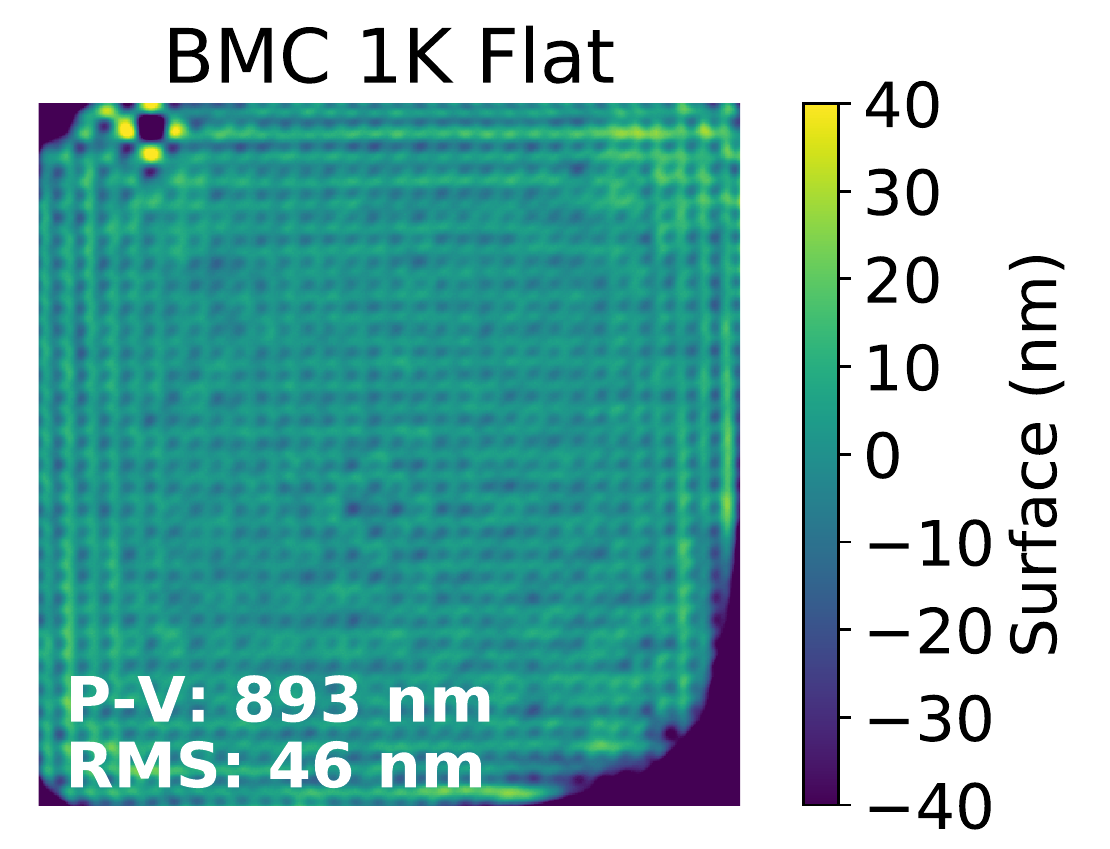}
   \includegraphics[height=6cm]{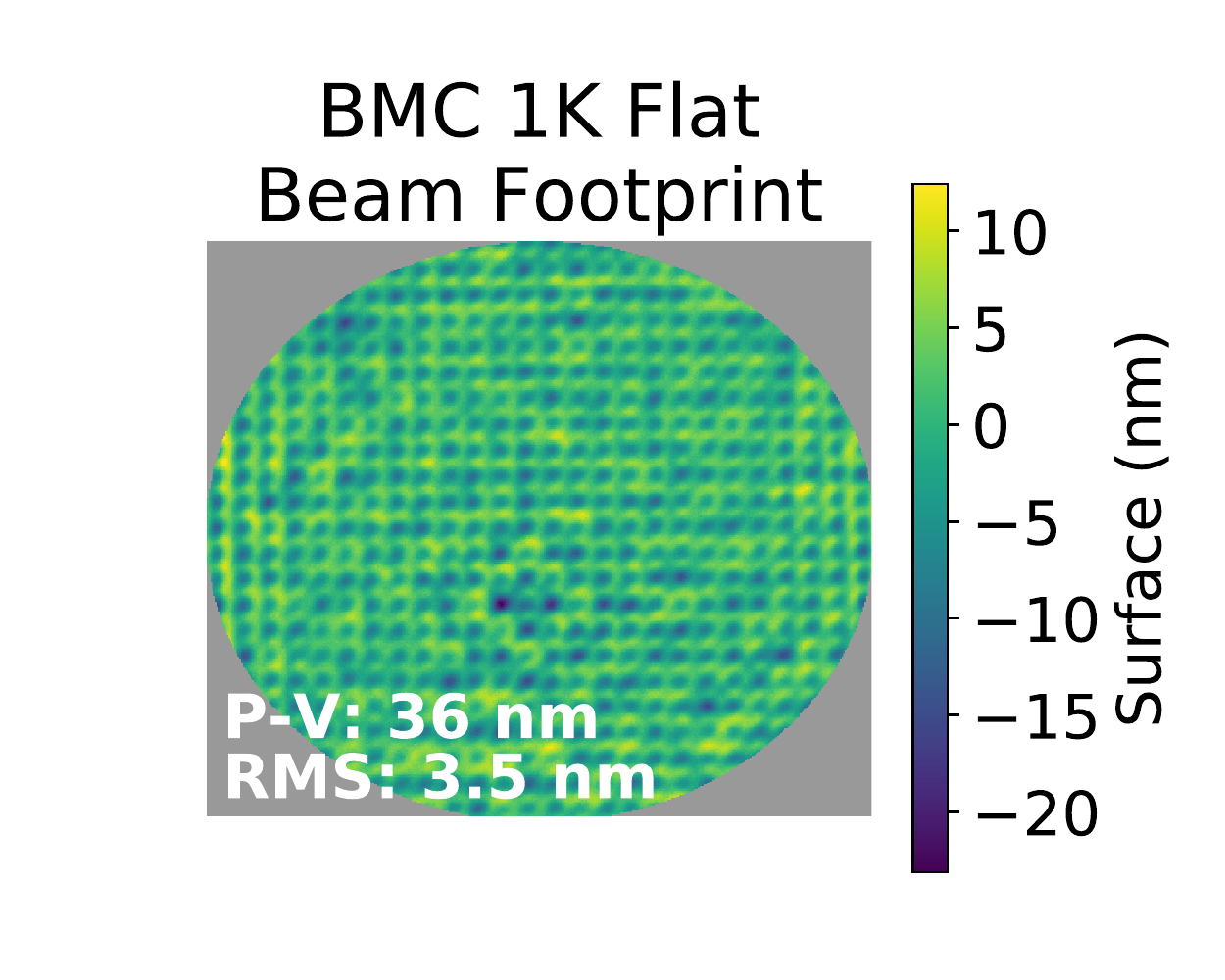}
   \end{tabular}
   \end{center}
   \caption[example]
   { \label{fig:bmc_1k} Left: Best flat produced in closed-loop with the Zygo interferometer over the entire surface of a Kilo-DM. A large uncorrectable slope in the lower right significantly increased the RMS. Right: The same flat limited to the beam footprint on the UA Wavefront Control Testbed.}
\end{figure}

With a BMC Kilo-DM (32x32) currently in use in the UAWFC lab, we have demonstrated the ability to characterize and flatten a similar DM following the procedure described in Sec. \ref{sec:pipeline}. In closed loop with the Zygo Verifire, we used an influence matrix derived from lab measurements to drive the Kilo-DM to a powered flat of 46nm over the entire surface and 3.5nm RMS over our beam footprint (Fig. \ref{fig:bmc_1k}).

\subsection{ALPAO DM97}
\label{sec:alpao} 

\begin{figure}[b]
   \begin{center}
   \begin{tabular}{c c}
   \includegraphics[height=3cm]{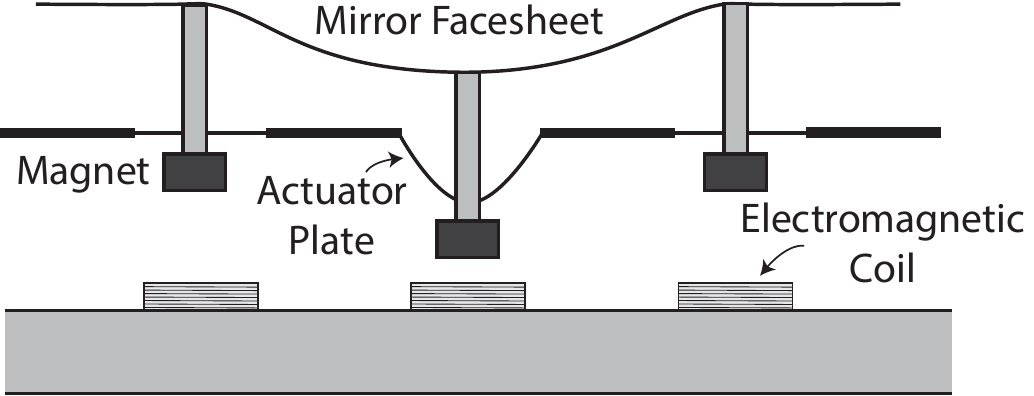}
   \raisebox{-0.1\height}{\includegraphics[height=4cm]{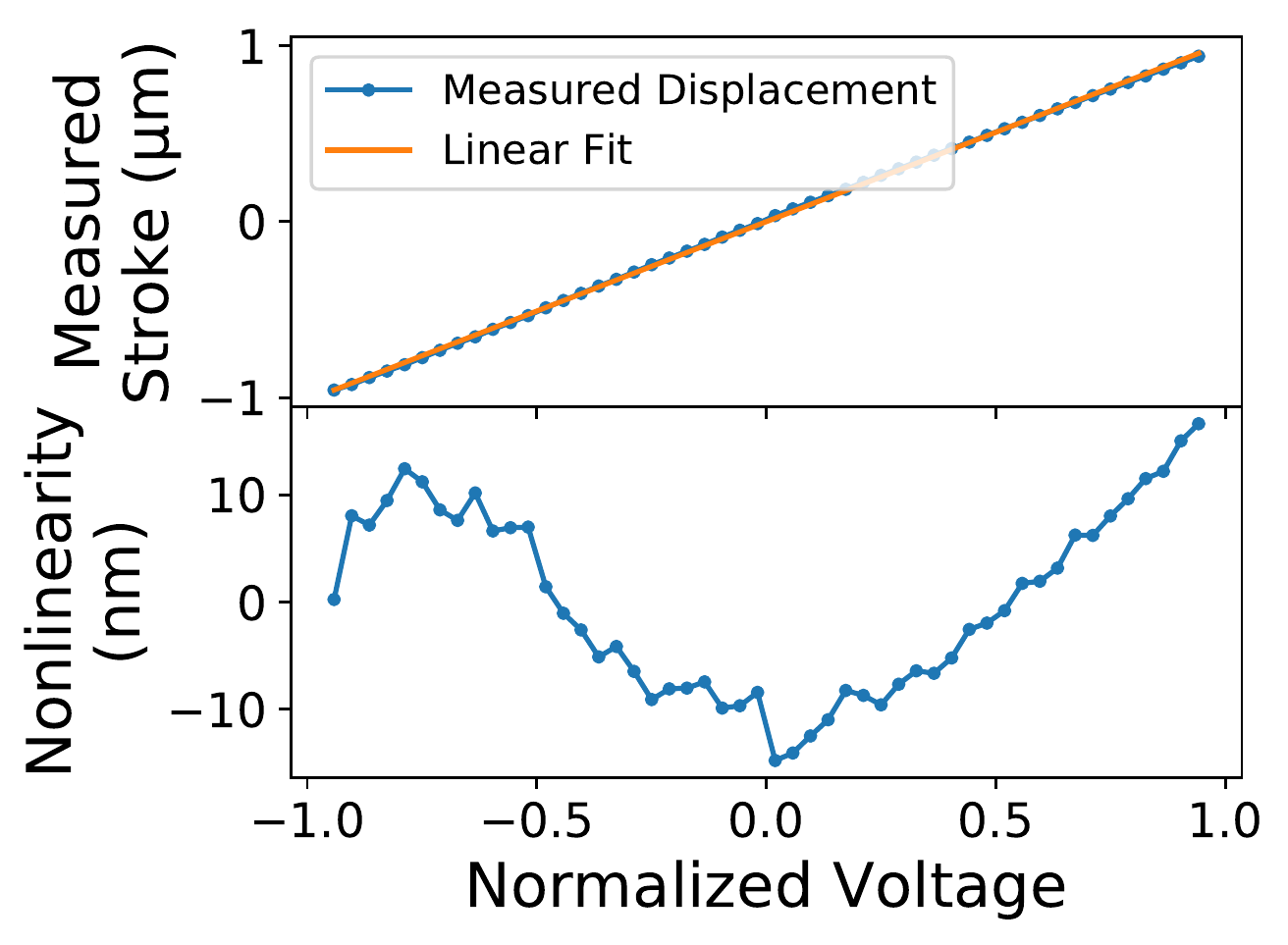}}
   \end{tabular}
   \end{center}
   \caption[example]
   { \label{fig:alpao} Left: A schematic of the ALPAO DM97. Current applied to the electromagnetic coils moves the reflective facesheet up or down. Right: The stroke of a single actuator has an approximately linear relationship with the applied voltage.}
\end{figure}

\begin{figure}[t]
   \begin{center}
   \begin{tabular}{c c}
   \includegraphics[height=6cm]{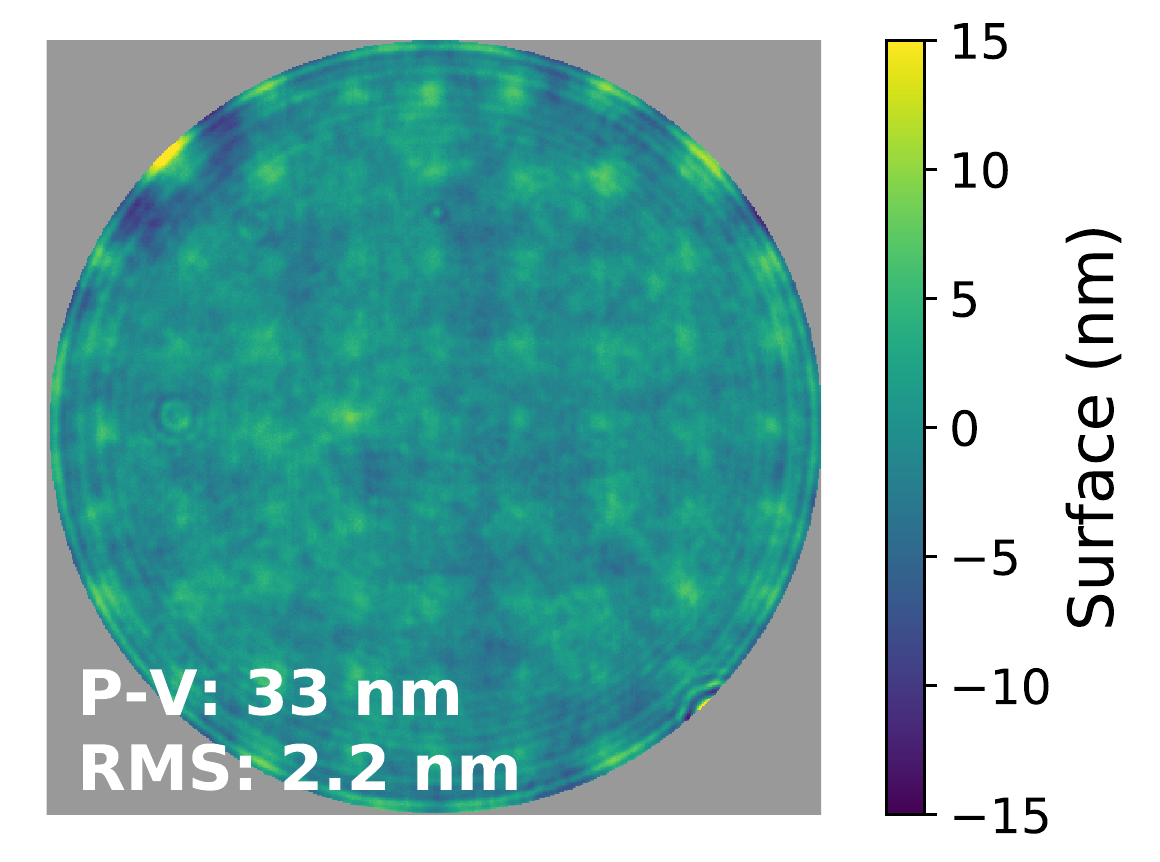}
   \includegraphics[height=6cm]{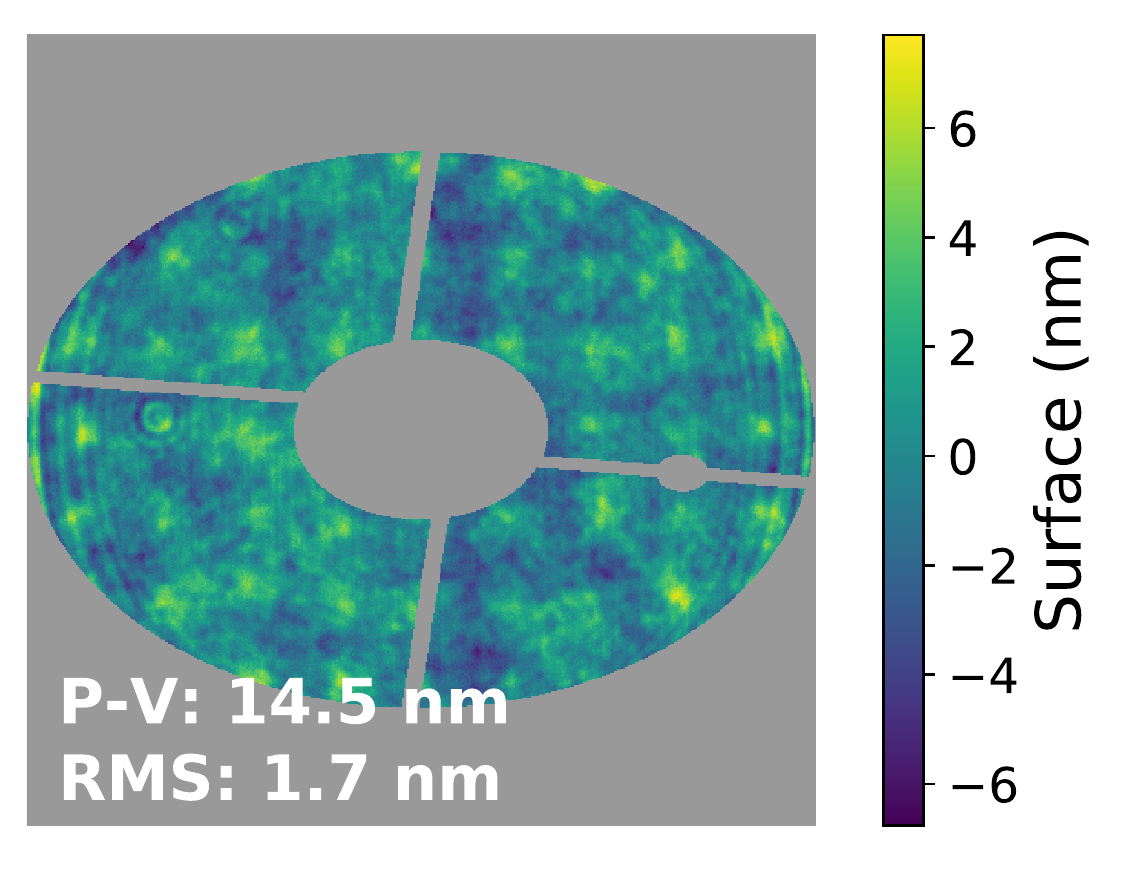}
   \end{tabular}
   \end{center}
   \caption[example]
   { \label{fig:alpao_surface} Left: The best flat over the full aperture of the ALPAO DM97. Right: The ALPAO flat with the MagAO-X pupil projected onto it. The beam is incident on the DM at a $45^\circ$ angle, creating the elongated pupil seen above.}
\end{figure}

The ALPAO DM97 (11x11) is a compact electromagnetic deformable mirror. It actuates via an electromagnetic coil and permanent magnet mounted by post to a reflective facesheet, resulting in an approximately linear voltage-stroke relationship over a wide range of strokes (Fig. \ref{fig:alpao}). The specifications of the device slated for low-order wavefront correction on MagAO-X are:
\begin{itemize}
\item 97 actuators
\item 13.5mm circular aperture
\item 30$\mu$m wavefront tip/tilt stroke
\item 1.5mm pitch
\end{itemize}
Following our characterization process, we measured each influence function and constructed a command matrix from them. In closed loop with the Zygo, we converged in approximately 3 steps to a flat of 2.2nm RMS over the full 13.5mm diameter and 1.7nm RMS over the MagAO-X pupil projected onto the NCP DM in the coronagraph arm (Fig. \ref{fig:alpao_surface}).

\section{FUTURE WORK}
\label{sec:thefuture}

DM characterization efforts will continue until integration into the MagAO-X instrument in fall 2018. In particular, we will focus on:

\begin{itemize}
\item improving surface metrology by use of a high-reflectance Zygo reference flat and the removal of the pellicle from the optical path.
\item performing in-house characterization of the BMC 2K and a second ALPAO DM97 upon delivery.
\item developing a DM model that accounts for effect of the nonlinear stroke response that arises from inter-actuator coupling, which becomes important for predictive control schemes and probe-based focal-plane wavefront sensing.
\end{itemize}

\acknowledgments      
 
This work was supported (in part) by NSF MRI Award \#1625441 \textit{Development of a Visible Wavelength Extreme Adaptive Optics Coronagraphic Imager for the 6.5 meter Magellan Telescope} (PI: Jared Males).



\bibliography{report}   
\bibliographystyle{spiebib}   

\end{document}